\documentclass[letterpaper,notoc,12pt]{JHEP}   
\usepackage{amssymb}
\usepackage[dvips]{epsfig}
\newcommand{\be}{\begin{equation}}
\newcommand{\ee}{\end{equation}}
\newcommand{\bea}{\begin{eqnarray}}
\newcommand{\eea}{\end{eqnarray}}
\newcommand{\nn}{\nonumber}

\newcommand{\spa}{\ \ \ }
\newcommand{\eqref}[1]{(\ref{#1})}

\newcommand{\sun}{\circledcirc}
\newcommand{\figpath}[1]{fig/#1}
\renewcommand{\baselinestretch}{1.2}

    \title{A New Non-Commutative Field Theory}
    \author{Konstantin Savvidy\\
    {\it Perimeter Institute, 35 King St N, Waterloo Ontario, N2J 2W9, Canada}\\
    E-mail: \email{konstantin@PerimeterInstitute.ca}
    }

    \abstract{
In this note we investigate a new type of non-commutative field theory
based on a constant skew-symmetric three-form parameter.  In $3+1$
dimensions such a three-form parameter can be viewed as a
short-distance regulator which nevertheless preserves spatial-rotation
and at long range preserves Lorentz invariance approximately.  For a
scalar field theory with quartic self-interaction we obtain
drastically improved ultra-violet behavior of the diagrams, due to
the oscillatory dependence of the interaction vertex on the momenta.  The
radiative corrections to the coupling are rendered finite already at the
one-loop level.
The key finding of this paper is that what appears as the reemergence
of UV divergences as IR singularity in $p \rightarrow 0$ limit, must
be interpreted simply as the logarithmic running of the coupling.
Thus at low energies the theory is virtually indistinguishable from
the standard theory.  Conversely at high energies the diagram
converges exponentially fast, the running of the coupling stops and
the theory avoids developing the Landau pole.  Bare coupling defined
at high energy can be kept small, and in this sense the theory is
similar to asymptotically free theories.  }

\preprint{\hepth{0205292}}


\begin{document}

\section{Introduction}
\label{secintro}
\renewcommand{\baselinestretch}{0.8}
Non-commutative theories have been extensively studied recently in
connection with string theory \cite{CDS,SW}.  Non-commutative geometry
and field theory naturally arises as a limit of string theory in the
presence of the anti-symmetric two-form NS-NS field $B_{\mu\nu}$.
Such field theories have been also studied in their own right with an
emphasis on the possibility of viewing the non-commutativity as a
short-distance regulator \cite{Filk:1996dm, Krajewski:1999ja, Varilly:1998gq}.  
Studies of perturbative dynamics have revealed the so-called
UV-IR connection \cite{Minwalla:1999px, VanRaamsdonk:2000rr},
\emph{i.e.} the strong IR effects appearing due to integrating out the
high momentum modes.

The author was originally motivated by the
intriguing problem of finding a generalization of non-commutative
field theory in the case that the B-field is linear, \emph{i.e.} its field
strength $H=dB$ is constant.  Unfortunately the string sigma model is
no longer soluble exactly  in this case and so far little progress has
been made attacking the problem directly.

We propose in Section \ref{secsun} an ad-hoc construct for a non-local field
theory with a three-form skew-symmetric parameter.  Formally the development
follows closely that of the conventional non-commutative theory, but in
several important physical aspects the results deviate significantly.

Renormalizability of the theory is checked up to two loop order in
Sections \ref{sec1loop}, \ref{sec2loop}.  The four-point (coupling
renormalization) amplitude is finite already at the one-loop order.  We
conjecture that the four-point function diagrams are finite at
arbitrary loop-order after taking into account the renormalization of
divergent two-point subdiagrams.

In conventional non-commutative theory the planar diagrams get overall
phases \cite{Filk:1996dm} that do not depend on the internal
momenta, the loop integrals being exactly the same as in the
corresponding diagrams of the commutative theory.  This results in
non-local counterterms which nonetheless are of the same form as the
original lagrangian.  By contrast, in the present theory the divergent
parts of the diagram do not depend on the quantum parameter $\theta$, thus
the necessary counterterms are local.  These arise only for the case
when the terms in the original lagrangian were local as well.

An important property of the theory is that the diagrams which are
rendered finite by the non-commutative cutoff are nonetheless
singular as the spatial part of external momentum tends to zero.
We interpret this in terms of the conventional renormalization
group approach as the logarithmic running of the coupling
constant. The relation between the bare coupling at high momenta
and the apparent physical coupling at low momenta is the
same as in conventional $\phi^{4}$ field theory.
At high energies convergence of the diagram is exponential, so
corrections to the bare coupling are small.  This means that unlike
conventional field theory the running of the coupling eventually stops,
allowing to keep the bare coupling small and avoiding the Landau pole.

\section{The $\sun$-Product}
\label{secsun}
We would like to construct a quantum space equipped with an anti-symmetric
three-form fuzziness parameter.  Define a non-local $\sun$-product%
\footnote{Pronounced as ``sun-product''.} of three functions, while leaving the
binary product intact,
\be
( f \sun g \sun h ) \, \vert_{x_0} = e^{ i
\, \theta_{\mu\nu\rho} \frac{\partial }{\partial x_\mu} \frac{\partial
}{\partial y_\nu} \frac{\partial }{\partial z_\rho}}~~ f(x) \, g(y) \,
h(z) ~\vert_{x=y=z=x_0}~~~.
\label{sun}
\ee
Here $\theta_{\mu\nu\rho}$ is a completely anti-symmetric three-form in $d$ dimensions,
however we restrict ourselves to the case $d=4$ throughout the paper.
In four spacetime dimensions such a form has exactly four independent
components: $\theta_{123}, \theta_{012}, \theta_{013}, \theta_{023}$.
Provided that the vector dual to our three-form
is time-like, \emph{i.e.} $\theta_{123}^2 > \theta_{012}^2 + \theta_{013}^2 + \theta_{023}^2$,
one can always choose a coordinate system in which only $\theta_{123}$ is nonzero.
Thus without loss of generality we set $\theta_{ijk}= \theta \,
\epsilon_{ijk}$ -- implying the absence of time-like components and
consequently time derivatives in the definition of the sun-product
\eqref{sun}.  This will nevertheless allow for a construction of a
Galilean theory: spatial rotation and translation invariance are
preserved.

The $\theta$ parameter has mass dimension $[-3]$, therefore we
introduce a convenient mass scale for the fuzziness of space
$\mu=\theta^{-1/3}$.  Naively, experiments in the infrared $p \ll \mu$
should not be able to discern the quantum structure of the underlying
space, because the sun-product \eqref{sun} goes over to ordinary
product in this limit.  In reality non-commutative field theory
exhibits strong IR effects
\cite{Minwalla:1999px,VanRaamsdonk:2000rr,Grimstrup:2002nr},
and we find that the present theory is no exception.

We note also that the underlying quantum structure of the space is not
clear as yet, in the sense of the equivalence between the Weyl
operator ordering on the quantum space and the Moyal star-product on
the commutative space.  It is tempting to speculate that the algebra of
the coordinates may be obtained by considering the
sun-product of the coordinates themselves.
In this way we arrive at
the following anomalous Jacobi-like identity, 
\be
x_{1} [x_{2}, x_{3}] + x_{2} [x_{3}, x_{1}] + x_{3} [x_{1}, x_{2}] = i \, \theta
\label{jacobi}
\ee
We do not make any use of this and leave the investigation of this quantum space to the future.
In this paper we concentrate instead on the perturbative properties of the quantum field theory
on this space.

\section{The Interaction Vertex}
\label{secvertex}
Consider the quantum field theory of one real scalar with
quartic self-interaction:
\be
{\cal L} = \int d^4 x \, \left[ \frac{1}{2} \, \partial_\mu \phi \,
\partial^\mu \phi  - \frac{1}{2} \, m^2 \phi^2 - \frac{\lambda}{4!} \,
(\phi \sun \phi \sun \phi ) \, \phi \right] ~~.
\ee
It is clear that the kinetic and mass terms are not modified, for the
same reasons as in the standard non-commutative theory.  However, the
interaction term is understood as a non-local momentum-dependent
four-point vertex (Figure \ref{fig0}). Had we attempted a different prescription
for the interaction from the one above, it would be reduced to the commutative case.
For example, $\phi \sun \phi \sun (\phi^2)$ is equivalent to $\phi^4$ due to conservation of momentum.

In momentum representation the interaction vertex is
obtained by symmetrization with respect to permutations of the
external legs
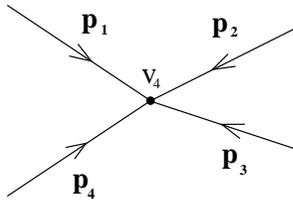
\FIGURE[h]{
\epsfbox{\figpath{four_vertex.eps}}
\caption{
The basic four-point vertex of the theory.
}
\label{fig0}
}
\be
V_{(4)} = \frac{\lambda}{4!} \sum_{a \neq b\neq c} \exp ( i \, \theta \, \epsilon_{ijk} \,
p^a_i \, p^b_j \, p^c_k ) =
\lambda \cos( \theta  \, p^1_{ [i } \, p^2_j \, p^3_{ k] }  ) ~~.
\label{vertex}
\ee

\begin{figure}[h]
\centerline{\epsfbox{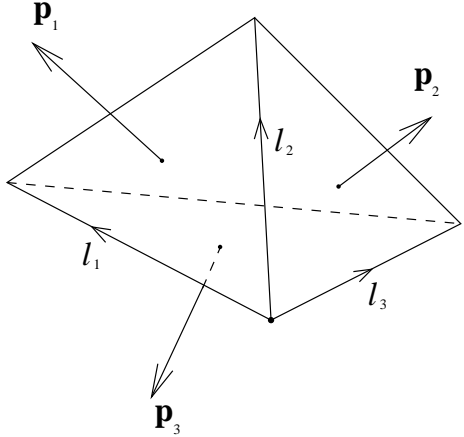}}
\caption{Given four arbitrary vectors that add up to nil,
one can always construct a tetrahedron with the four vectors being
the area-normals of the faces, $\vec{p}_1=\frac 1 2 \, \vec{l}_1\times\vec{l}_2$
\emph{etc}...
Compute:
$8 \, \vec{p}_1\cdot(\vec{p}_2\times\vec{p}_3)=
[ \vec{l}_1\times\vec{l}_2 ] \cdot \left[ [ \vec{l}_2\times\vec{l}_3
] \times [\vec{l}_3\times\vec{l}_1 ]\right]=
 \left(\vec{l}_2\cdot[\vec{l}_3\times\vec{l}_1]\right)^2 $,
and this quantity  is equal to the volume of the tetrahedron squared.}
\label{fig1}
\end{figure}

\noindent
Because of momentum conservation all twenty-four terms above
arising from reordering have the same magnitude, twelve occurring
with plus sign and twelve with a minus, for example
$p^1_{[i} \, p^2_j \, p^3_{k]} = - p^2_{[i} \, p^3_j \, p^4_{k]}$. Moreover, it
is possible to give the following geometrical interpretation of
this phase (Figure \ref{fig1}).  Construct a tetrahedron with the
four (spatial) momenta being the vectors normal to the faces of
the tetrahedron and equal to the area of the corresponding face.
The phase in~\eqref{vertex} is equal exactly to the square of the
volume of the tetrahedron (see caption to Figure \ref{fig1}).

The non-locality and non-renormalizability  introduced through the higher-derivative
interactions both in standard non-commutative theory and in our sun-product theory
is potentially troublesome, in that the $\theta$-expanded interactions are non-renormalizable
but the very special exponential structure prevents the appearance of most problems
generically associated with such non-renormalizable interactions.

Note that $\vert V^{(4)} \vert \leq \lambda$ : the vertex is bounded
from above by its standard value.  Therefore, the convergence of all
diagrams is improved, or at worst unchanged.
This may enable us to use the standard theorems about renormalizability
at higher loop order and the disentanglement of overlapping
divergences \cite{Chepelev:1999tt}.  One must worry that the necessary
counterterms may not be of the same non-local form as in the original
lagrangian, however it turns out that only local counterterms for the
two-point function appear.  We illustrate these issues in Sections
\ref{sec1loop}, \ref{sec2loop} by computing the divergent and finite
parts of the one- and two-loop diagrams.


\section{One-loop Renormalization}
\label{sec1loop}
\begin{figure}
\centerline{\epsfbox{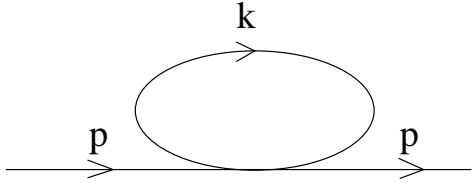}  }
\caption[fig2]{ The one-loop mass  renormalization diagram:
the sun-phase vanishes due to the collinearity of any three of the legs at the vertex.
Conventional regularization is necessary to cancel the quadratic divergence.}
\label{fig2}
\end{figure}

Let us see what happens in this theory with the one-loop
renormalization of mass. Since any three legs of the vertex in
the diagram (Figure~\ref{fig2})
have linearly dependent momenta, being $p, k, -p, -k$,  the
phase factor is trivial. This means that the diagram is still
quadratically divergent. Perhaps this is a good thing, as
the non-locality of the theory does not give rise to new
infrared singularities in the propagator (at least in this order)
\bea
i\,\Gamma^{(2)} = \lambda \, \int \frac{d^4k}{(2\pi)^4} \,
\frac{1}{k^2+m^2}=
\frac \lambda {32\pi^2} \, \left( \Lambda^2  -  m^2 \,   \ln \frac {\Lambda^2} {m^2} \right)
\eea
\begin{figure}
\centerline{\hbox{\epsfbox{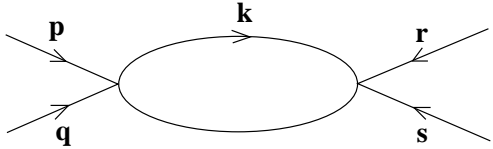}}}
\caption[fig3]{ The one-loop coupling  renormalization diagram:
the sun-phase does not vanish at either of the two vertices for generic values of momenta,
being $\mathbf {k \cdot(q \times p)}$ and $\mathbf {k \cdot(r \times s)}$ respectively.
The contribution of this diagram is finite.}
\label{fig3}
\end{figure}
Thus the counterterms necessary to cancel the divergence in this
one-loop mass renormalization diagram are indeed of the same form as
the lagrangian we started with.  It turns out that at two-loop order
(see Section \ref{sec2loop}) the mass renormalization diagram contains
both convergent and divergent contributions.  The divergent part is of
the same form as the original lagrangian and is local, while only the
finite part contains nontrivial dependence on the quantum parameter
$\theta$.

Presently we consider the one-loop radiative
correction to the coupling (Figure \ref{fig3}).  Momenta in bold
represent spatial vectors
%
\bea
i\,\Gamma^{(4)}=\frac{\lambda^2}{2} \,
\int \frac{dk_0 \, d^3\mathbf{k}}{(2\pi)^4} ~ \frac{ \cos
\theta \, \mathbf {k \cdot(p \times q)} } { k^2+m^2 }
\,
\frac{ \cos{ \theta \, \mathbf {r \cdot(s \times k)} } } {
(p+q+k)^2+m^2 }       \hspace{3.9cm} \nn \\
%
= \frac{\lambda^2}{32 \pi^2}\int_0^{\infty} d\alpha d\beta ~(\alpha+\beta)^{-2} e^{-( \alpha
+ \beta )m^2} e^{-\frac{\alpha\beta}{\alpha+\beta}(p+q)^2}
e^{-\frac{\theta^2}{\alpha+\beta}(\mathbf{p} \times \mathbf{q} +
\mathbf{r} \times \mathbf{s})^2}~+ [\mathbf{r} \leftrightarrows \mathbf{s}]
\eea
Upon introducing the convenient variables $z = \alpha+\beta$ and $u = \alpha/z$, obtain
\be
i\,\Gamma^{(4)} = \frac{\lambda^2}{32 \pi^2}
\int_0^{\infty} z^{-1} \, dz
  \int_{0}^{+1} du~
    e^{-z m^2} \,
    e^{ -z u(1-u)(p+q)^2 }   \,
    e^{-\frac{\theta^2}{z}( \mathbf{p} \times \mathbf{q} +
\mathbf{r} \times \mathbf{s} )^2} ~+ [\mathbf{r} \leftrightarrows \mathbf{s}]~~.
\label{zint}
\ee
This is now identical to the standard result, with the expression
\mbox{$\Lambda_{eff}^{-2} = \theta^2 (\mathbf{p} \times \mathbf{q}\pm \mathbf{r} \times \mathbf{s})^2$}
playing the role of the cutoff length scale. The integral in
\eqref{zint} would have been logarithmically
divergent near $z \rightarrow 0$ without this cutoff.
Thus, for sufficiently large $\Lambda_{eff}^{2}$, the $z$ integral can be approximated
by its leading logarithmic divergence,
\be
i\,\Gamma^{(4)} = \frac{\lambda^2}{32 \pi^2} \,
\left[\ln \frac{(\mathbf{p} \times \mathbf{q} + \mathbf{r} \times
\mathbf{s})^2}{\mu^{4}} + [\mathbf{r} \leftrightarrows \mathbf{s}]+
\int_{0}^{1} du ~\ln \frac{m^2+(p+q)^2 u(1-u) }{\mu^{2}}
~\right]
\label{4ptresult}
\ee
For small values of $\Lambda_{eff}^{2}$ the integral converges much
faster, in fact exponentially.

At this point we recognize the reemergence of the UV-IR mixing of \cite{Minwalla:1999px, VanRaamsdonk:2000rr}.
The amplitude \eqref{4ptresult} is divergent for some accidental values of non-zero momenta,
but more importantly it is singular as $\mathbf{p} \rightarrow 0$. In Section \ref{effaction}
we show that this does not lead to genuine IR singularity because the effective action remains finite.

In conventional field theory the logarithmic divergence in the four-point function
is interpreted in terms of renormalization of the coupling \cite{Callan:1975vs, Gross:1975vu}.
We would like to make
the connection between our theory and the standard one. For that, we should consider the
first term in \eqref{4ptresult} as the correction to the effective coupling at low energy.
Taking into account the additional contributions in the $\mathbf{t,u}$ channels,
\be
\lambda_{eff} = \lambda - \frac{3 \lambda^{2}}{8\pi^{2}} \, \ln \frac{|\mathbf{p}|}{\mu}
\textrm{~~~for small~~} \mathbf{p}
\ee
For large $\mathbf{p}$ the behavior of the integral is different, and $\lambda_{eff}$ is exponentially
close to $\lambda$. In this way, the running of the coupling stops once momenta reach
the fuzziness scale $\mathbf{p} >> \mu$. Thus in this theory there is no Landau pole singularity.
However, there is instead the possibility of destabilization at nonperturbatively low momenta
$|\mathbf{p}| \sim \mu \exp (-1/\lambda)$. We fully expect that higher loop diagrams will remove this apparent
pathology, changing the behavior at such low momenta to
\be
\lambda_{eff} =\frac{ \lambda  }{1 + \frac{3 \lambda}{8\pi^{2}} \, \ln \frac{|\mathbf{p}|}{\mu}}
\ee
as is the modification of the one-loop result
inferred from standard renormalization group approach \cite{Callan:1975vs, Gross:1975vu}.


\section{Two-loop Mass Renormalization}
\label{sec2loop}
Two-loop renormalization in non-commutative field theory was
considered in detail in \cite{Aref'eva:1999sn, Micu:2000xj, Huang:2000ka}.  We
consider the two-loop order primarily because at one-loop order the
two-point function does not receive non-trivial $\theta$-dependent
corrections.

\begin{figure}[h]
\centerline{\hbox{\epsfbox{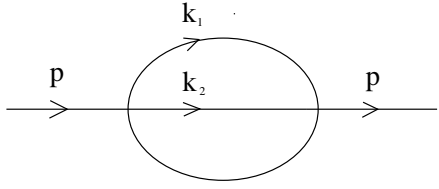}}}
\caption[fig4]{ Two-loop mass renormalization diagram. The vertex phase is nontrivial, but
                is the same for both vertices. This leads to both a finite and an infinite contribution.
                Since the infinite contribution is $\theta$ independent, the infinity can be cancelled by
                a local counterterm.  }
\label{fig4}
\end{figure}

In order to obtain radiative corrections to the mass at two loop order (Figure \ref{fig4})
we should compute the double integral
\be
\lambda^2 \int\int \frac{d^4k_1}{(2\pi)^4} \, \frac{d^4k_2}{(2\pi)^4}  ~
\frac{ \cos\mathbf{\theta \, p \cdot (k_1 \times k_2})  }{
(k_1^2+m^2) (k_2^2+m^2)} \, \frac{\cos\mathbf{\theta \,p \cdot (k_1 \times k_2) }}{( p+k_1 +k_2)^2+m^2}~~.
\ee
This integral has overall quadratic divergence by power counting,
which is not completely cured by the appearance of the phases.  The
product of the two cosines has a zero-frequency component as well as a
$2 \, \theta \,\mathbf{p \cdot (k_1 \times k_2)}$ component.  Such zero-frequency components do not seem
to persist at higher loop order. We therefore conjecture that the two-point
function at three and higher loop order are finite after the renormalization of
divergent one- and two- loop subdiagrams.

The finite part of the integral can be computed as usual with the introduction
of Schwinger parameters $\alpha, \beta, \gamma$ and performing the
gaussian integrations over momenta
\be {\displaystyle \int _{0}^{\infty }}
{\displaystyle \frac { e^{ - (\alpha+\beta+\gamma) m^2} \, e^{ - \frac
{\alpha \beta \gamma} {\alpha\beta+\beta\gamma+\alpha\gamma}\,p^2 } }
{
(\alpha\beta+\beta\gamma+\alpha\gamma)\,(\alpha\beta+\beta\gamma+\alpha\gamma
+ \theta ^2\,\mathbf{p}^2 ) }} \,d\alpha\,d\beta\,d\gamma = \frac {
\mathcal{F}( m^2 \theta |\mathbf{p}| , p^2 \theta |\mathbf{p}|
)}{\theta |\mathbf{p}|}
\ee
where $\mathcal{F}$ is a function weakly
dependent on its dimensionless arguments.
%
%

Again, the result is analogous to the standard one, if we identify
$\theta |\mathbf{p}|$ with the effective cutoff.
Aside from the leading $1/\theta |\mathbf{p}|$ behaviour there is
also subleading term $\sim p^{2} \ln p^{2} \theta |\mathbf{p}|$
corresponding to the necessary
wavefunction renormalization at two loops in ordinary field theory.
Again, the divergence of this diagram,
which is normally manifest in the ultra-violet region, here reappears in the infrared:
$ \mathbf{p} \rightarrow 0$.

\section{Effective Action}
\label{effaction}
The interaction of UV and IR in non-commutative field theories leads to
troublesome IR singularities.  It has been suggested that these IR
effects can be explained through the appearance of closed string modes
in string theory.  Another possibility might have been to introduce
extra IR cutoffs.  In \cite{Grimstrup:2002nr} a suggestion was made to
solve this problem through a non-local field redefinition, however the
field transformation is potentially singular, and in any case the
problem reappears at the second loop order.  Therefore we are
compelled to investigate this issue in the present theory.

The appropriate tool is to check whether the effective action is finite,
after the removal of the UV cutoff $\Lambda$.  The quadratic one-loop
effective action in our theory is
\be
\Gamma^{(2)}_{eff} = \int d^4p
\, \phi(p) \, \phi(-p) \, \left[ p^2 + m^2 + \frac \lambda {32\pi^2}
\, \left( \Lambda^2 - m^2 \ln \frac {\Lambda^2} {m^2} \right) \right]
\ee
The divergent part as $\Lambda \rightarrow \infty$ is removed as usual by
reabsorbing into the definition of the mass.

The one-loop effective four-point function is read off from \eqref{4ptresult}
\bea
i\,\Gamma^{(4)}_{eff} = \frac{\lambda^2}{32
\pi^2} \, \int d^4p_1 \, d^4p_2 \, d^4p_3 \, d^4p_4 \spa \phi(p_1) \,
\phi(p_2) \, \phi(p_3) \, \phi(p_4) ~ \delta^4(p_1+ p_2+ p_3+ p_4)
\nn\\
\left[ \ln \frac{(\mathbf{p_1} \times \mathbf{p_2} + \mathbf{p_3} \times
\mathbf{p_4})^2}{\mu^{4}}  +
\int_{0}^{1} d\alpha ~ \ln \frac{m^2+(p_1+p_2)^2 \alpha(1-\alpha)}{\mu^{2}}\right]~~~
\eea
Despite the fact that there is a logarithmic divergence in the integrand
near $\mathbf{p} \rightarrow 0$, this is more than compensated by the measure in the integral.
The IR singularity in the two-loop two-point function is potentially more troublesome,
since it is not a logarithmic one as above, but is a pole as $ \mathbf{p} \rightarrow 0$.
\be
\Gamma^{(2)}_{two~loop} = \int dp_o \, d^3\mathbf{p} \, \phi(p) \, \phi(-p) \,
\left[ p^2 + M^2 + \lambda^2  \frac { \mathcal{F}( m^2 \theta |\mathbf{p}| , p^2 \theta |\mathbf{p}| )}{\theta |\mathbf{p}|} \right]
\ee
The crucial difference with standard non-commutative theory is that the effective cutoff
appearing above involves momenta along all three spatial directions,
while in star-product theory only momenta in the non-commutative plane appear.
Thus, the integral above is actually convergent near $\mathbf{p} \rightarrow 0$.
The theory is free of the IR singularity.

\section*{Acknowledgments}
I would like to thank  J. Ambj\o rn, L.~Alvarez-Gaume, J.~Brodie, L.~Freidel,
R.~Myers, L.~Smolin, N.~Seiberg and R.J.~Szabo for
extensive discussions and useful comments.

\addcontentsline{toc}{section}{References}
\renewcommand{\baselinestretch}{1}


\bibliographystyle{JHEP}

%

\bibliography{noncomm}

\providecommand{\href}[2]{#2}\begingroup\raggedright\begin{thebibliography}{10}

\bibitem{CDS}
A.~Connes, M.~R. Douglas, and A.~Schwarz, {\it Noncommutative geometry and
  matrix theory: Compactification on tori},  {\em JHEP} {\bf 02} (1998) 003,
  [\href{http://xxx.lanl.gov/abs/http://arXiv.org/abs/hep-th/9711162}{{\tt
  http://arXiv.org/abs/hep-th/9711162}}].

\bibitem{SW}
N.~Seiberg and E.~Witten, {\it String theory and noncommutative geometry},
  {\em JHEP} {\bf 09} (1999) 032,
  [\href{http://xxx.lanl.gov/abs/http://arXiv.org/abs/hep-th/9908142}{{\tt
  http://arXiv.org/abs/hep-th/9908142}}].

\bibitem{Filk:1996dm}
T.~Filk, {\it Divergencies in a field theory on quantum space},  {\em Phys.
  Lett.} {\bf B376} (1996) 53--58.

\bibitem{Krajewski:1999ja}
T.~Krajewski and R.~Wulkenhaar, {\it Perturbative quantum gauge fields on the
  noncommutative torus},  {\em Int. J. Mod. Phys.} {\bf A15} (2000) 1011--1030,
  [\href{http://xxx.lanl.gov/abs/http://arXiv.org/abs/hep-th/9903187}{{\tt
  http://arXiv.org/abs/hep-th/9903187}}].

\bibitem{Varilly:1998gq}
J.~C. Varilly and J.~M. Gracia-Bondia, {\it On the ultraviolet behaviour of
  quantum fields over noncommutative manifolds},  {\em Int. J. Mod. Phys.} {\bf
  A14} (1999) 1305,
  [\href{http://xxx.lanl.gov/abs/http://arXiv.org/abs/hep-th/9804001}{{\tt
  http://arXiv.org/abs/hep-th/9804001}}].

\bibitem{Minwalla:1999px}
S.~Minwalla, M.~Van~Raamsdonk, and N.~Seiberg, {\it Noncommutative perturbative
  dynamics},  {\em JHEP} {\bf 02} (2000) 020,
  [\href{http://xxx.lanl.gov/abs/http://arXiv.org/abs/hep-th/9912072}{{\tt
  http://arXiv.org/abs/hep-th/9912072}}].

\bibitem{VanRaamsdonk:2000rr}
M.~Van~Raamsdonk and N.~Seiberg, {\it Comments on noncommutative perturbative
  dynamics},  {\em JHEP} {\bf 03} (2000) 035,
  [\href{http://xxx.lanl.gov/abs/http://arXiv.org/abs/hep-th/0002186}{{\tt
  http://arXiv.org/abs/hep-th/0002186}}].

\bibitem{Grimstrup:2002nr}
J.~M. Grimstrup, H.~Grosse, L.~Popp, V.~Putz, M.~Schweda, M.~Wickenhauser, and
  R.~Wulkenhaar, {\it Ir-singularities in noncommutative perturbative
  dynamics},
  \href{http://xxx.lanl.gov/abs/http://arXiv.org/abs/hep-th/0202093}{{\tt
  http://arXiv.org/abs/hep-th/0202093}}.

\bibitem{Chepelev:1999tt}
I.~Chepelev and R.~Roiban, {\it Renormalization of quantum field theories on
  noncommutative r**d. i: Scalars},  {\em JHEP} {\bf 05} (2000) 037,
  [\href{http://xxx.lanl.gov/abs/http://arXiv.org/abs/hep-th/9911098}{{\tt
  http://arXiv.org/abs/hep-th/9911098}}].

  \bibitem{Callan:1975vs}
C.~G. Callan, {\it Introduction to renormalization theory}, In *Les Houches
  1975, Proceedings, Methods In Field Theory*, Amsterdam 1976, 41-77.

\bibitem{Gross:1975vu}
D.~J. Gross, {\it Applications of the renormalization group to high-energy
  physics}, In *Les Houches 1975, Proceedings, Methods In Field Theory*,
  Amsterdam 1976, 141-250.

\bibitem{Aref'eva:1999sn}
I.~Y. Aref'eva, D.~M. Belov, and A.~S. Koshelev, {\it Two-loop diagrams in
  noncommutative phi**4(4) theory},  {\em Phys. Lett.} {\bf B476} (2000)
  431--436,
  [\href{http://xxx.lanl.gov/abs/http://arXiv.org/abs/hep-th/9912075}{{\tt
  http://arXiv.org/abs/hep-th/9912075}}].

\bibitem{Micu:2000xj}
A.~Micu and M.~M. Sheikh~Jabbari, {\it Noncommutative phi**4 theory at two
  loops},  {\em JHEP} {\bf 01} (2001) 025,
  [\href{http://xxx.lanl.gov/abs/http://arXiv.org/abs/hep-th/0008057}{{\tt
  http://arXiv.org/abs/hep-th/0008057}}].

  \bibitem{Huang:2000ka}
W.-H. Huang, {\it Two-loop effective potential in noncommutative scalar field
  theory},  {\em Phys. Lett.} {\bf B496} (2000) 206--211,
  [\href{http://xxx.lanl.gov/abs/http://arXiv.org/abs/hep-th/0009067}{{\tt
  http://arXiv.org/abs/hep-th/0009067}}].

\end{thebibliography}\endgroup
\raggedright

\end{document}